\documentstyle[12pt,twoside,psfig,fleqn,espcrc1]{article}

\newcommand{\Dslash}{\rlap{\,/}D}

\newcommand{\cond}{\langle \bar{\psi}{\psi}\rangle}
\newcommand{\dbary}{\langle n \rangle}
\newcommand{\Tr}{{\rm tr}}

\newcommand{\AmS}{{\protect\the\textfont2
  A\kern-.1667em\lower.5ex\hbox{M}\kern-.125emS}}

\title{Random Matrices and the Glasgow Method}

\author{Mikl\'{o}s \'{A}d\'{a}m Hal\'{a}sz
         \address{Department of Physics and Astronomy, 
                  State University of New York at Stony Brook, \\ 
                  Stony Brook, NY 11794-3800, USA}
}

\begin{document}
\maketitle

\begin{abstract}
A simple non-Hermitean random matrix (RM) model is used to 
study the Glasgow method of finite-density lattice QCD. 
The zeros of the RM partition function are evaluated through an
averaging procedure, involving the zeros of the random 'propagator matrix' in the 
complex chemical-potential plane. 

The nature of the uncertainty affecting the results is similar to that 
produced by rounding errors in computing the known analytic result. 
This similarity is exploited to give quantitative estimates on the 
relationship between the size of the matrix and the number of configurations 
needed to achieve a given precision. 
For the quenched ensemble considered here, the relationship is exponential.

\end{abstract}

\section{INTRODUCTION}

Random matrix theory has had a significant impact on the theoretical and
lattice study of QCD \cite{Verb98}. This is especially true for 
lattice QCD at finite chemical potential \cite{Step96}.
RM models are solvable, easy to simulate, and reproduce
some of the difficulties of lattice calculations.

Lattice simulations at finite chemical potential $\mu$ are plagued by the 
so-called sign problem.
The hermiticity of the $\mu$-dependent piece of the Dirac operator
$\Dslash = \Dslash_{\mu=0} + \mu \gamma_0 $
is opposite to that of the $\mu=0$ part. This is due to  
 the Euclidean formulation. 
As a result, the fermion determinant is in general 
a complex number and cannot be used as a probability measure. Therefore, 
generating an unquenched ensemble of configurations at nonzero chemical 
potential is not possible. 

The quenched approximation has been shown to fail in the context of the 
random matrix model \cite{Step96,Hala96,Hala97} employed here. 
Quenched simulations probe not QCD but the $N_f \rightarrow 0$ limit of a 
theory where the fermion determinant is replaced by its absolute value 
\cite{Step96}. 
One may simulate the unquenched theory at nonzero $\mu$,
using an ensemble generated under different conditions,
and including the fermion determinant in the observable.  
An elegant way to do this is provided by the Glasgow method \cite{Barb98}, 
with recent promising results \cite{Barb97}.

The unquenched QCD partition function at arbitrary $\mu$ can be seen as the 
expectation value of the fermion determinant, averaged over an ensemble 
weighted only with the gauge part of the action. One may also use an
unquenched ensemble at zero chemical potential, but then the observable will
also include the inverse fermion determinant at $\mu=0$:
\begin{equation}
Z ~=~ \left\langle \det \left( \Dslash(\mu) \right) \right\rangle_{gauge}
~=~ \left\langle  \left. \det \left( \Dslash(\mu)\right) \right/
  \det \left(\Dslash(0)\right) \right\rangle_{gauge,fermi,\mu=0}
\end{equation}
The efficiency of the averaging process is  determined
by the overlap between the region of phase space probed by one's ensemble
and that of the 'true', $\mu \ne 0$ ensemble. 
The unquenched ensemble at $\mu=0$ should do better in this respect than the 
quenched one. 
\footnote{It should be emphasized, though, that both approaches are in 
principle correct. 
Using the quenched ensemble is not equivalent to the 
quenched approximation, where the determinant is neglected altogether. } 

The idea of the Glasgow method is to expand the fermion determinant,
 which is an $N \times N$ matrix on the lattice, in powers of the 
fugacity $\xi=e^\mu$ ($\mu$ enters in this combination). 
The expansion is finite and exact.
It is obtained by 
rewriting the fermion determinant as
$\det \left(\Dslash(\mu)\right) = \xi^{N} \det \left( P + \xi \right) $
where $P$ is called the propagator matrix. 
The eigenvalues of $P$
can be used to construct the expansion of $\det \Dslash$.

Since $\xi$ is the same for all 
configurations, the averaged coefficients give the expansion of the partition 
function in powers of $\xi$.  Thus, with only one $\mu$-independent ensemble,
one has access to the full $\mu$-dependence of the partition function. 
The location of the zeros in the complex $\mu$ plane maps out the phase 
structure.
In particular, the zeros of $Z(\mu)$ close to the real axis define the critical
value(s) of the chemical potential \cite{LeeY52}.

\section{RANDOM MATRIX MODEL}

We consider a model defined by the partition function 
\cite{Step96,Hala96,Hala97,Hala98}
\begin{equation}
\label{model}
Z(m, \mu) = \int {\cal D} C e^{- N \Tr C C^\dagger} \det ( D(m,\mu) )~~~;
~~~D(m,\mu) ~=~ \left[ \begin{array}{cc} 
               m & i C + \mu \\ i C^\dagger + \mu & m \end{array} \right]
\end{equation}
where $C$ is an $N \times N$ complex matrix, and $\int{\cal D} C$ stands for 
integration over each complex matrix element. 
This defines an ensemble of random matrices, which mimics the ensemble
of gauge field configurations from QCD.
The exponential factor ensures that the real and imaginary parts of each 
matrix element of $C$ are independent Gaussian random numbers. 
$D(m,\mu)$ approximates the Dirac operator for quark mass $m$. 
It is anti-Hermitean for $\mu=0$, 
with no definite hermiticity otherwise.
The chiral condensate and the number density are the logarithmic derivative
of (\ref{model}) with respect to $m$ and $\mu$, respectively.

\subsection{The random propagator matrix and its eigenvalues}

Define the random analog, $P(m)$ of the fermion propagator matrix as follows:
\begin{equation}
\label{fmatrix}
P(m) ~=~ \left( \begin{array}{cc}
 i C & m   \\ m  & i C^\dagger 
\end{array} \right) ~~\rightarrow~~
\det ( D (m,\mu )) ~=~ \det ( P(m)  + \mu ) .
\end{equation}
Its main property is that its eigenvalues are the roots
of the fermion determinant. Compute the baryon number density 
by substituting (\ref{fmatrix}) into (\ref{model}) and taking the logarithmic
derivative. The result is similar to the formula for the condensate using
the Dirac matrix.
\begin{equation}
\label{barydens}
\dbary = \partial_\mu \ln Z = 
\left\langle \Tr \left( \frac{1}{P(m) + \mu} \right) \right\rangle_Z ~~;~~
\cond = \partial_m \ln Z = 
\left\langle \Tr \left( \frac{1}{D(\mu, 0 ) + m} \right) \right\rangle_Z .
\end{equation}
Here $\langle \cdots \rangle_Z $ stands for averaging with the partition 
function (\ref{model}).
The propagator matrix is not Hermitean in general, so its
eigenvalues are scattered in the complex plane. For $m=0$, they are uniformly
distributed in the unit disc.
Their distribution can be computed using the phase-quenched partition function,
the same way it is done for the Dirac operator \cite{Step96}.

\subsection{Glasgow averaging}

Consider an ensemble of $N \times N$ matrices $C$, generated with the Gaussian
weight in (\ref{model}). For each $C$, we construct the corresponding fermion 
matrix and compute its eigenvalues. This way the fermion determinant in 
(\ref{model}) can be written as a polynomial in $\mu$,
\begin{equation}
\label{polyconfig}
\det ( D(m, \mu)) ~=~ \det (P(m) + \mu) ~=~
 \prod\limits_{k}(\lambda_k + \mu) ~=~ \sum\limits_{k=1}^{2N} c_k \mu^k .
\end{equation}
The average over the Gaussian ensemble performed on (\ref{polyconfig}) will 
commute with the finite sum. One obtains the partition function (\ref{model})
as a polynomial in $\mu$:
\begin{equation}
\label{polypart}
Z(m,\mu) ~=~ \langle \det (D(m,\mu)) \rangle_{Gauss} 
~=~ \sum\limits_{k=1}^{2N}  \langle c_k \rangle_{Gauss} \mu^k 
~=~ \prod\limits_{k=1}^{2N} (\Lambda_k + \mu) .
\end{equation}
Here, the $\Lambda_k$ stand for the zeros of the partition function. They
may be interpreted as 'average eigenvalues' of the propagator matrix. 

\subsection{Exact zeros of the partition function}

The zeros of the partition function (\ref{model}) are  accessible
by direct calculation \cite{Hala96,Hala97}. 
The exact zeros for $m=0$ and $N=96$ are identical to those in 
the last panel of Fig.\ref{many3}.
The zeros are located along the critical line 
$ Re( 1 + \mu^2 + \ln \mu ) =0 $. 
It can be derived from the formula, correct to lowest order in $1/N$,
\begin{equation}
\label{twopieces}
Z (m , \mu) = \sqrt{ \frac{2 \pi^3}{N}} e^{- N (1 + \mu^2)} ~+~
\frac{\pi}{( 1 + \mu^2) N} ~ \mu^{2 (N +1 )} .
\end{equation}
In the thermodynamic limit $N \rightarrow \infty$, the partition function 
(\ref{model}) splits naturally into two terms, given by simple analytic 
functions of the parameters. Each of them dominates exponentially
in some region of parameter space, corresponding to a thermodynamic phase. The 
derivatives of $Z$ are discontinuous on the boundary separating these 
regions. In the thermodynamic limit, the zeros coalesce into a cut that 
follows this boundary. 
The critical value of $\mu_c = 0.527$ (for $m=0$) is determined by the point where the 
line crosses the real axis.

\section{SIMULATIONS}

\subsection{Some results}

\begin{figure}
\vspace{-0.1cm}
\centerline{
\hspace{-0.5cm}
\psfig{file=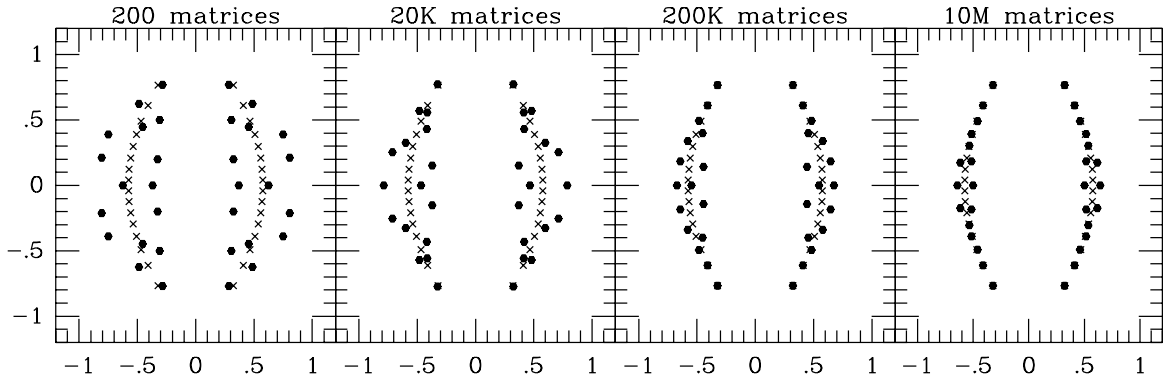,height=4.9cm,angle=0}}
\vspace{-0.7cm}
\caption{Different stages of Glasgow averaging, for $m=0$ and $N=16$.  }
\label{many_short}
\vspace{-0.5cm}
\end{figure}

Results of Glasgow averaging for $N=16$ and $m=0$ are shown in 
Fig.\ref{many_short} .
Complex $N \times N$ matrices $C$ are generated with Gaussian weight. 
The eigenvalues of the corresponding propagator matrix are calculated. 
The polynomial whose roots are the eigenvalues is constructed for each
'configuration'. The coefficients are then averaged. The points shown in the
figure are the roots of the averaged polynomial for a certain number of 
configurations.
For comparison, the location of the exact zeros is indicated with crosses.

The roots for a few configurations are very similar to quenched eigenvalues, 
since the eigenvalues of individual configurations are controlled by the
quenched distribution.
There is a clear indication that the result tends to 
evolve into the true zeros. 
There are two interesting points about how convergence is approached:
(i) the cloud of zeros, even far from convergence, has a fairly regular shape; 
(ii) some zeros converge much faster than others. 
In our case, the zeros with the largest imaginary parts converge
first. 

\subsection{Artificial noise}

Since the coefficients are obtained through averaging, they are 
affected by a random uncertainty, similar to a rounding error. 
It is resonable to assume that the \em 
relative error \em is of the same order on each coefficient. 
One can study the effect of such an uncertainty by adding a fixed proportion
of noise to the exact coefficients:
$\tilde{c}_k = c_k (1 + \epsilon R_k )$ ,
where $\epsilon$ is a fixed small number and $R_k$ are uniform real random
numbers between $\pm 1$.

\begin{figure}
\centerline{
\hspace{-0.5cm}
\psfig{file=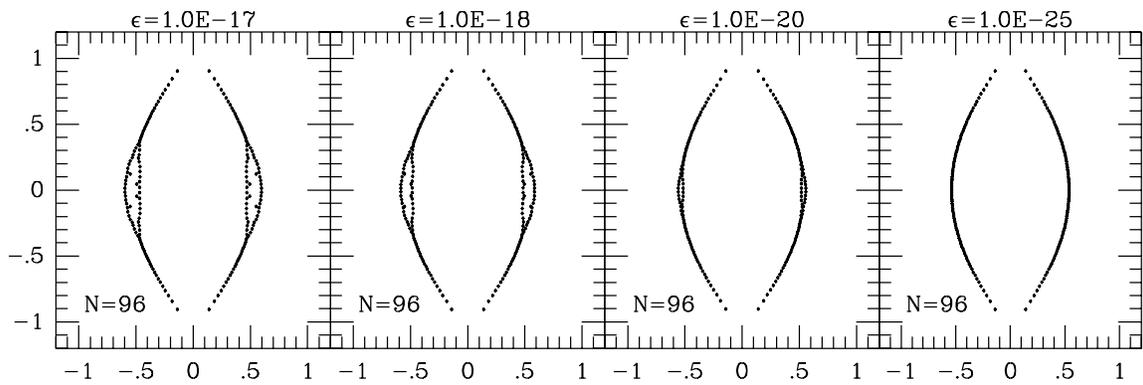,height=4.9cm,angle=0}}
\vspace{-0.7cm}
\caption{Effect of artificial noise on the exact zeros of the partition 
function. 
}
\label{many3}
\vspace{-0.5cm}
\end{figure}

The effect of artificial noise is dramatic as seen in Fig.\ref{many3} .
A noise factor of the order of $10^{-17}$ alters significantly the value of 
$\mu_c$ . 
These diagrams are strikingly similar to the 
last two in Fig.\ref{many_short} . 
Again, the zeros closest to the real axis are the most sensitive.

It is reasonable to expect that the effect of the errors is an exponential
function of the matrix size $N$. For artificial noise, we can afford to check 
this.
\begin{figure}
\centerline{
\hspace{-0.5cm}
\psfig{file=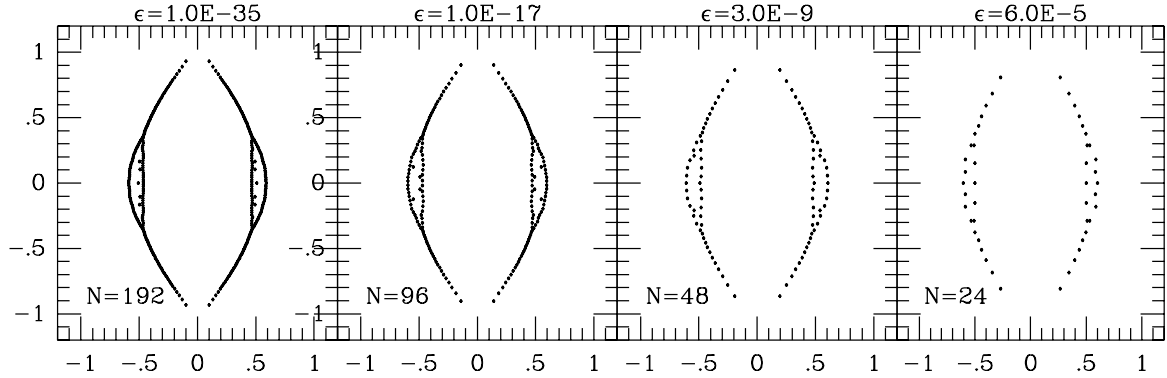,height=4.9cm,angle=0}}
\vspace{-0.6cm}
\caption{Artificial noise for different matrix sizes, tuned to lead to similar
distortion .
}
\label{many2}
\vspace{-0.4cm}
\end{figure}
In Fig.\ref{many2} the noise factor $\epsilon$ has been 'tuned' to have  
approximately the same effect 
for each matrix size $N$, which is doubled from one frame to the next.
The corresponding noise factor $\epsilon$ is roughly the square of the previous
one. Indeed, the artificial noise factor that has the same effect for different
matrix sizes is proportional to a small number to the power $N$ : 
$\epsilon_N \sim e^{-\alpha N}$ . In other words, the precision in digits 
required to determine $\mu_c$ with given accuracy, is proportional to the 
matrix size.

\subsection{Estimations and speculations}

Our assumptions need to be checked before making statements about simulations.
Is the relative variance $\sigma(c_k)/c_k$ of the coefficients always the same?
From the existing results, one can say that for given $N$, the
relative variance is fairly constant. For example, in a very long
simulation for $N=16$, it is very close to $4.00$, except for the first few
coefficients, for which it is smaller with a factor of $2..3$. For larger
values of $N$, the relative variance tends in general to be constant within
one set, and approximately given by $\log_2 N$ .

An important question refers to the correlations between coefficients. For
instace, if all coefficients fluctuate proportionally, the effect on the 
roots of
the polynomial is smaller than if the fluctuations are uncorrelated. At this
point, there is no conclusive information on correlations. Assuming that
they are negligible, we  estimate the number of Glasgow configurations 
needed to get some fixed accuracy on $\mu_c$ , as a function of matrix size 
$N$, 
\begin{equation}
\label{estimate}
\frac{\log_2 N}{N_{conf}^{1/2}} ~=~ \frac{\Delta c_k}{c_k} ~=~
 \epsilon ~=~ e^{- \alpha N} ~~\rightarrow~~ 
N_{conf} \approx (\log_2 N)^2 ~ e^{2 \alpha N} .
\end{equation}
This estimate is hard to check, given the large numbers
involved. Consider $N=16$ and $N=8$. We have
$\epsilon_{16} = \epsilon_{8}^2 \rightarrow 
\left( 4 / (N_{conf}^{(16)})^{1/2} \right) =
\left( 3 / (N_{conf}^{(8)})^{1/2} \right)^2 \rightarrow$ 
$N_{conf}^{(8)} \approx 2 \sqrt{ N_{conf}^{(16)}} $.

\begin{figure}
\centerline{
\hspace{-0.5cm}
\psfig{file=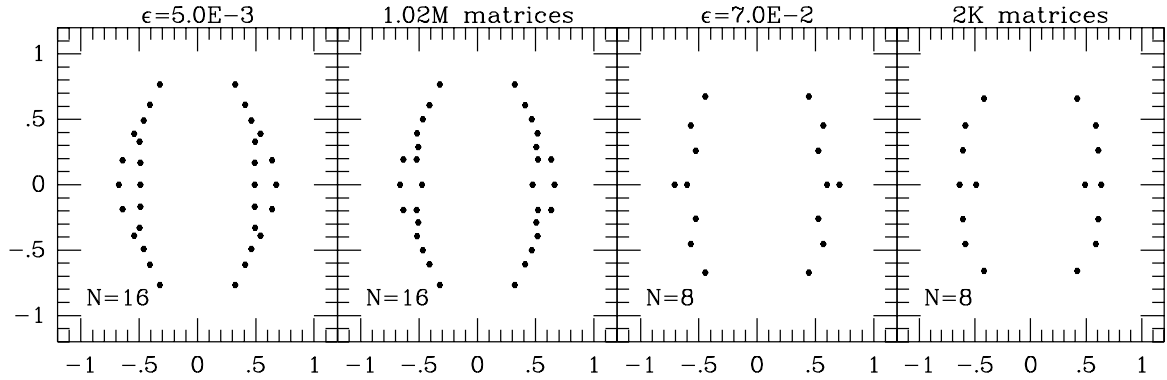,height=4.9cm,angle=0}}
\vspace{-0.6cm}
\caption{ Simulations and artificial noise, so that the precision
of $\mu_c$ is the same.
}
\label{many4}
\vspace{-0.4cm}
\end{figure}

The comparison for artificial noise with $\epsilon_8=0.07$ and 
$\epsilon_{16}=0.005$, and simulations with $2000$,
respectively one million configurations is made in Fig.\ref{many4} .
Our estimations seem justified as all four panels have 
approximately the same error on $\mu_c$ . For $N=256$, which in terms of
lattice size would correspond to a $4^4$ lattice with one degree of freedom
per site, one gets $\epsilon_{256} \sim 10^{-40}$ leading to a number of
configurations around $10^{80}$ .

\section{CONCLUSIONS}

In the context of a schematic random matrix model, it is found that
the number of configurations needed to achieve a fixed precision using
the Glasgow method, grows exponentially with the matrix size $N$. 
This is not surprising, since the ensemble of matrices used for the 
calculation is quenched. 
As a result, there is probably very little overlap between the region of 
'phase space' probed by this ensemble and the one important in the 
unquenched ensemble at nonzero chemical potential. 
In terms of convergence, the procedure used here is equivalent
to 'brute force' (using a quenched ensemble and including the complex 
determinant in the observable). 
The problem of achieving a large cancellation while averaging over a 
complex phase \cite{Hala97} is traded for the huge precision needed for the 
coefficients in order to obtain the zeros with reasonable accuracy. 

On the positive side, the Glasgow method is seen to converge here, albeit
for small $N$. 
The qualitative picture is slightly different from what
is implied in \cite{Barb97}. The cloud of partition function 
zeros slowly narrows 
down to the true critical line, which is not distinguishable as long as 
the cloud is present. 
Some zeros converge faster than others. In general, the points in parameter 
space where the continuum limit partition function is larger are less 
sensitive to perturbations, leading to more accurate results. The
\em limiting \em partition function is \em not \em zero along the critical 
line.
The zeros along the 'higher' sections of the critical line converge first.
 
An open question is to what extent an unquenched ensemble at
zero chemical potential, like in lattice simulations,
would improve the situation. 
The present model lacks certain symmetries that are present on the lattice.
They may play a significant role, and can be included in a more sophisticated
RM model.

\vspace{0.3cm}

\noindent{\bf Acknowledgements}

This work was supported by the US DOE grant DE-FG-88ER40388.
I thank my collaborators, Jac Verbaarschot and Misha Stephanov.  
I am grateful to I. Barbour, M. Lombardo, S. Morrison , S. Hands and 
A. Galante for  
helpful discussions, and especially to Ely Klepfish who suggested this 
problem. Thanks are due to F. Karsch and M. Lombardo, who
organized the Bielefeld workshop.


\begin{thebibliography}{9}
\bibitem{Verb98} J.J.M. Verbaarschot, these proceedings, and
 references therein.
\bibitem{Step96} M.A. Stephanov, Phys. Rev. Lett. 76 (1996) 4472 . 
\bibitem{Hala96} M.\'{A} Hal\'{a}sz, A.D. Jackson. J.J.M. Verbaarschot,
Phys. Lett. B 395 (1996) 293 .
\bibitem{Hala97} M.\'{A}. Hal\'{a}sz, A.D. Jackson. J.J.M. Verbaarschot,
Phys. Rev. D56 (1997) 5140 .
\bibitem{Barb98} see I.M. Barbour and S.E. Morrison, these proceedings, and
references therein.
\bibitem{Barb97} I.M. Barbour, S.E. Morrison, E.G. Klepfish, J.B. Kogut,
M.P. Lombardo, Phys. Rev. D56 (1997) 7063 .
\bibitem{LeeY52} C.N. Yang and T.D. Lee, Phys. Rev. 87 (1952) 404 and 410 .
\bibitem{Hala98} M.\'{A} Hal\'{a}sz, A.D. Jackson, R.E. Shrock, M.A. Stephanov,
A.D. Jackson, hep-ph/9804290 .
\end{thebibliography}
\end{document}